# Optimizing Higher-order Lagrangian Perturbation Theory for Cold Dark Matter Models

Arno G. Weiß

*Max-Planck-Institut für Astrophysik, Karl-Schwarzschild-Str. 1,
D–85748 Garching, F.R.G. (aow@mpa-garching.mpg.de)*

Stefan Gottlöber

*Astrophysikalisches Institut Potsdam, An der Sternwarte 16,
D–14482 Potsdam, F.R.G. (sgottloeber@aip.de)*

Thomas Buchert

*Theoretische Physik, Ludwig-Maximilians-Universität, Theresienstr. 37,
D–80333 München, F.R.G. (buchert@stat.physik.uni-muenchen.de)*

**Abstract.** We report on the performance of Lagrangian perturbation theory up to the second order for the standard cold dark matter (SCDM) and broken scale invariance (BSI) scenarios. We normalize both models to the COBE data, the BSI model serves as an example of models which fit the small-scale power of galaxy surveys. We optimize Lagrangian perturbation solutions by removing small-scale power from the initial data and compare the results with those of numerical simulations. We find an excellent performance of the optimized Lagrangian schemes down to scales around the correlation length or smaller, depending on the statistics used for the comparison. The optimization scheme can be expressed in a way which is independent of the type of fluctuation spectrum and of the size of the simulations.

## 1. Lagrangian Perturbation Theory and Optimization

For an in-depth discussion of the Lagrangian perturbation approach relevant to this work see (Buchert 1994), and for the optimization approach (Buchert et al. 1994, Melott et al. 1995, Weiß et al. 1995) and references therein.

Denoting comoving Eulerian coordinates by $\vec{q}$ and Lagrangian coordinates by $\vec{X}$, the field of trajectories $\vec{q} = \vec{F}(\vec{X}, t)$ up to the second order on an Einstein-de Sitter background reads:

$$\vec{F} = \vec{X} + q_1(a)\, \nabla_X \Psi^{(1)}(\vec{X}) + q_2(a)\, \nabla_X \Psi^{(2)}(\vec{X}), \qquad (1)$$

where the time-dependent coefficients can be expressed in terms of the expansion function $a(t) = (\frac{t}{t_i})^{2/3}$:

$$q_1 = \left(\frac{3}{2}\right)(a - 1), \qquad (2)$$

555

$$q_2 = \left(\frac{3}{2}\right)^2 \left(-\frac{3}{14}a^2 + \frac{3}{5}a - \frac{1}{2} + \frac{4}{35}a^{-\frac{3}{2}}\right). \tag{3}$$

The perturbation potentials have to be constructed by solving iteratively the two boundary value problems:

$$\Delta_X \Psi^{(1)} = I(\Psi_{,i,k})t_i, \tag{4}$$

$$\Delta_X \Psi^{(2)} = 2II(\Psi^{(1)}_{,i,k}), \tag{5}$$

where $I$ and $II$ denote the first and second principal scalar invariants of the tensor gradient ($\Psi^{(1)}_{,i,k}$). Under the restriction of parallelism of the initial peculiar-velocity field and the peculiar-acceleration field, we can set $\Psi^{(1)} = \Psi t_i$ ($\Psi$ being the initial peculiar-velocity potential), and the first-order flow-field (1) reduces to Zel'dovich's approximation, as discussed in (Buchert 1994) and references therein.

The optimization scheme used here for the Lagrangian perturbation theory was discussed in Coles et al. (1993) and further investigated for higher orders in Melott et al. (1995). The idea is to smooth away some of the small-scale fluctuations in the initial data. In a previous work, Melott et al. (1995), we found that convolution of the initial density field with a Gaussian window function $\hat{W}(k, k_{gs}) = \exp(-k^2/k_{gs}^2)$ of appropriate width $k_{gs}$ considerably improves on the performance of the Lagrangian perturbation schemes. Specifically, here we run a search-loop over different smoothing lengths applied to the initial data of the first and second order Lagrangian approximations. Then we compare the resulting density fields with those obtained with a PM-code (Kates et al. 1991) by determining the cross-correlation coefficient $S = \langle \delta_1 \delta_2 \rangle / \sigma_1 \sigma_2$ of the two fields (here, the mean $\langle \ldots \rangle$ is evaluated over all cells of the discrete density fields). As optimal we define that smoothing length $k_{gs}$ which maximizes $S$.

In a further analysis, we compute a scale-dependent cross-correlation $S(R_g)$ by smoothing the two final density fields $\delta_1$, $\delta_2$ with various filter widths $R_g$.

## 2. Numerical simulations with COBE-normalized SCDM and BSI initial data

In order to follow the nonlinear evolution of the formation of structure we have performed N-body simulations using a standard PM code (Kates et al. 1991) with $128^3$ particles on a $256^3$ grid (Kates et al. 1995). The universe is assumed to be spatially flat ($\Omega = 1$). It is dominated by cold dark matter. We consider here simulations which were performed in boxes of 500 $h^{-1}$ Mpc, 200 $h^{-1}$ Mpc and 75 $h^{-1}$ Mpc. The simulations were started with the power spectrum $P(k)$ of density perturbations calculated at $z = 25$,

$$P(k) = \left(\frac{\delta \rho}{\rho}\right)^2_{\vec{k}} = \frac{4}{9}\left(\frac{kR_h}{2}\right)^4 \Phi^2(k) T^2(k), \tag{6}$$

where $R_h = 2H^{-1}$ denotes the horizon scale. The primordial perturbation spectrum $\Phi$ is either the Harrison-Zel'dovich spectrum ($\Phi = const$) of the SCDM model or the spectrum with broken scale invariance calculated from a double inflationary model (Gottlöber et al. 1991). In the BSI model the primordial spectrum is of Harrison-Zel'dovich type both in the limit of very large and very small scales.

We have used the CDM transfer function $T(k)$ of Bond and Efstathiou (1984). For all simulations we have normalized our spectra using the 10°-variance of the CMB fluctuations $\sigma_T = (30 \pm 7.5)\mu K$ of the first year COBE data (Smoot et al. 1992). Thus, the power spectrum of the BSI model shows less power on small scales than the SCDM model. The scales of nonlinearity defined in eq.(7) below are $\lambda_{nl}^{SCDM} = 27h^{-1}$ Mpc and $\lambda_{nl}^{BSI} = 7h^{-1}$ Mpc at the time $z = 0$. In Fig. 1 we show the linear BSI and SCDM spectra and indicate the box sizes and the resolution of our simulations.

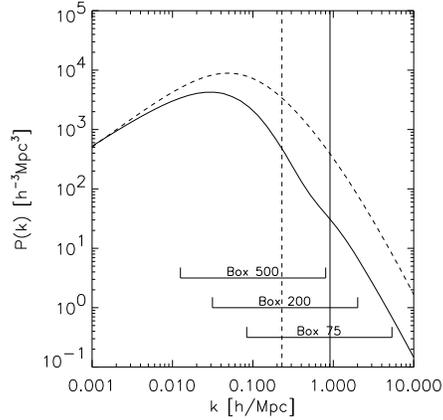

Figure 1. The linear power spectra of our SCDM (dashed line) and BSI (solid line) models at time $z = 0$.

The first and second year COBE data were analyzed by many authors using different statistical techniques. The new normalization of Górski et al. (1994) is about 25 % higher than the normalization which we used in the simulations. Consequently, in this normalization the scales of nonlinearity would increase to $\lambda_{nl}^{SCDM} = 33h^{-1}$ Mpc and $\lambda_{nl}^{BSI} = 11h^{-1}$ Mpc.

## 3. Results and Discussion

In an analysis of both models at different times ($z = 0, 1, 2$) we find that the optimal $k_{gs}$ can consistently be estimated from the scale of nonlinearity $k_{nl}$ of the considered spectrum and time evolution, $k_{nl}$ being defined by

$$\frac{a^2(t)}{(2\pi)^3} \int_0^{k_{nl}} d^3k \, P(k) = 1. \qquad (7)$$

As can be seen from Fig. 2, $k_{gs} \simeq 1.45 k_{nl}$ for the first-order and $k_{gs} \simeq 1.2 k_{nl}$ for the second-order Lagrangian perturbation solutions. It can be seen from the lines in Fig. 2, which are regression fits to the scatter plot of $k_{gs}/k_{nl}$ versus $n(k_{nl})$, that the optimal smoothing length does practically not depend on the slope $n(k_{nl})$ of the power spectrum at the nonlinearity scale, and is thus independent of the exact shape of the power spectrum. While the smoothing length for first-order shows some random scatter with $n(k_{nl})$, the optimal smoothing of the second order is much more robust against variations in the initial power spectrum.

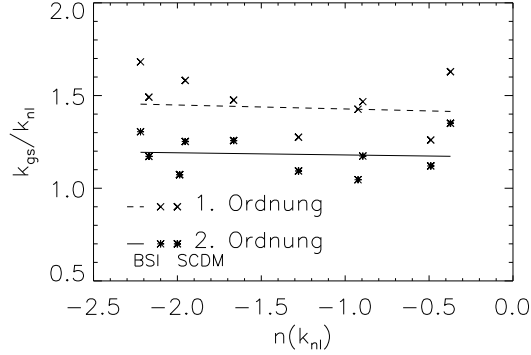

Figure 2. The optimal smoothing length $k_{gs}$ in units of $k_{nl}$ versus the slope $n(k_{nl})$ of the power spectrum at $k_{nl}$ for the BSI and SCDM models.

The scale-dependent cross-correlation function $S(R_g)$ measures primarily whether mass is moved to the right place. Fig. 3 shows its value for the comparison of the density fields at time $z = 0$. The performance of the optimized Lagrangian perturbation schemes is very good down to scales of $1 \ldots 2 h^{-1}$Mpc for both the SCDM and BSI models. This quality can clearly not be met by the Eulerian linear theory. (For this comparison we use an improvement on the Eulerian linear theory as proposed in Coles et al. (1993), called "chopped linear theory"; it complies to both $\delta_{clin} \geq -1$ and $\langle \delta_{clin} \rangle = 0$, even at late times. In terms of the linearly evolved $\delta_{lin}$ we set $1 + \delta_{clin} = \alpha(1 + \delta_{lin})$ if $\delta_{lin} > -1$ and 0 otherwise, where $\alpha$ is a normalization constant keeping the total mass the same.)

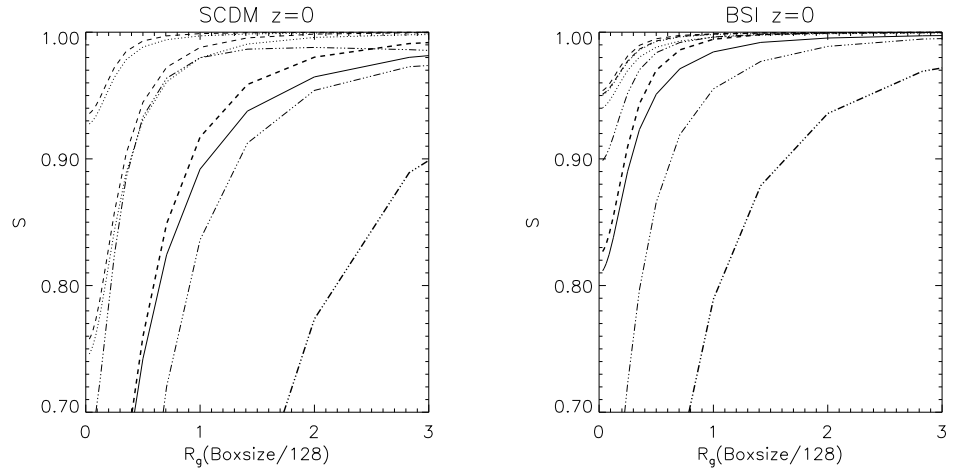

Figure 3. Scale-dependent cross-correlation function $S(R_g)$ of first-order (dash-dotted lines) and second-order (dashed lines) optimized Lagrangian perturbation theory as well as of "chopped linear theory" (dash-dot-dot-dotted lines) correlated with the results of the numerical simulations for the SCDM (left) and BSI (right) models, evolved to $z = 0$. Thin lines are for the $500 h^{-1}$Mpc boxes, medium thick lines for $200 h^{-1}$Mpc and thick lines for the $75 h^{-1}$Mpc boxes, respectively.

The two-point correlation function $\xi(r)$ of the numerical simulations is reproduced well by the optimized Lagrangian perturbation schemes down to about the "correlation length" $r_0$, $\xi(r_0) = 1$ (Fig. 4). Below $r_0$, which itself is underestimated by about 10%, the amplitude of $\xi(r)$ of the optimized Lagrangian perturbation schemes drops well below that of the numerical simulations. Still, the use of our optimization scheme leads to a considerable improvement in $\xi(r)$ on small scales, especially for models with a large amplitude of small-scale fluctuations, like the SCDM model.

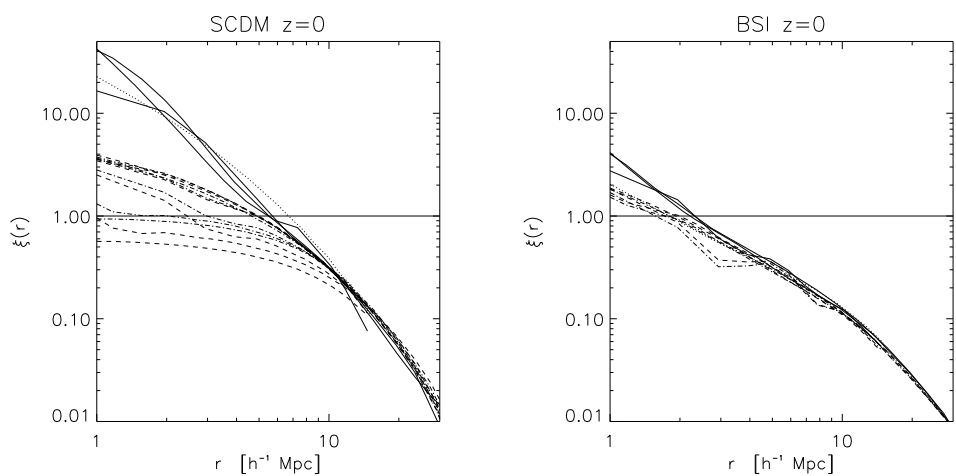

Figure 4. Two-point correlation function for the SDCM (left) and BSI (right) models at $z = 0$. Thick solid lines are for the numerical simulations, the dotted line is the prediction of the Eulerian linear theory, the thick (thin) dash-dotted and dashed lines are for first- and second-order optimized (unoptimized) Lagrangian perturbation solutions, respectively.

The optimized Lagrangian perturbation schemes reproduce the density field very good down to the scales of groups and clusters of galaxies, even deep into the nonlinear regime (i.e. today). Their inherent advantage is their high speed of execution, their complexity is comparable to one time step of a conventional particle-mesh N-body code. Thus, they allow the quick production of a large ensemble of simulations, e.g. for the evaluation of different cosmological models or for doing statistics over an ensemble of different realizations of a single model. Furthermore, the analyticity of the mapping (1) allows not only the local determination of the resulting shear stresses, e.g. for the introduction of local biasing schemes, it can also be used to enhance the particle density by interpolation of the mapping (1), so that one can use statistics with a high selection effect, like the simulation of galaxy catalogues with geometry and luminosity selection, even for very large-scale realizations (on Gpc-scales). The analysis of the singularity structure of the mapping (1) in order to identify building blocks of large-scale structure offers a way to analytically relate initial data to present-day structure (Buchert et al. 1995).

**Acknowledgments.** We wish to thank Adrian L. Melott (University of Kansas) for valuable discussions as well as for the permission to use his programs for the cross–correlation statistics, which were also used in Buchert et al. (1994) and Melott et al. (1995). AGW wishes to thank the AIP in Potsdam for the opportunity to work on this project during a stay at the AIP. TB acknowledges support of the "Sonderforschungsbereich **375** für Astro–Teilchenphysik der Deutschen Forschungsgemeinschaft". SG wishes to thank the MPA in Garching for its hospitality. AGW and TB like to thank the University of Valencia for hospitality during a working visit, which was financially supported by "acciones integradas", project AI95-14.